\documentclass[twoside]{article}
\usepackage[latin1]{inputenc}
\usepackage{t1enc}
\usepackage{a4}
\usepackage{tabularx}
\usepackage{epsf,psfig}

\def\bref{\vspace{4pt}\noindent\hangindent=10mm}
\def\arcmin{\ifmmode ^{\prime}\else$^{\prime}$\fi}
\def\arcsec{\ifmmode ^{\prime\prime}\else$^{\prime\prime}$\fi}
\def\approxlt{\mathrel{\hbox{\rlap{\lower.55ex \hbox {$\sim$}}
        \kern-.3em \raise.4ex \hbox{$<$}}}}
\def\approxgt{\mathrel{\hbox{\rlap{\lower.55ex \hbox {$\sim$}}
        \kern-.3em \raise.4ex \hbox{$>$}}}}

\begin{document}

\setcounter{figure}{0}
\setcounter{section}{0}
\setcounter{equation}{0}

\begin{center}
{\Large\bf
X-ray Evidence for Supermassive Black Holes at the Centers of Nearby, Non-Active Galaxies}\\[0.7cm]

Stefanie Komossa\\[0.17cm]
Max-Planck-Institut f\"ur extraterrestrische Physik \\
Postfach 1312, 85741 Garching, Germany \\
skomossa@mpe.mpg.de, http://www.xray.mpe.mpg.de/$\sim$skomossa/
\end{center}

\vspace{0.5cm}

\begin{abstract}
\noindent{\it

We first present a short overview 
of X-ray probes of the black hole region of active
galaxies (AGN) and then concentrate on the X-ray search for 
supermassive black holes (SMBHs) in optically
non-active galaxies.

The first part focuses on recent results from the X-ray observatories 
{\sl Chandra} and {\sl XMM-Newton} which detected a wealth of
new spectral features which originate in the nuclear region of AGN.  

In the last few years, giant-amplitude, non-recurrent X-ray flares have been
observed from several {\em non}-active galaxies.
All of them share similar
properties, namely:
extreme X-ray softness in outburst, huge peak luminosity (up to $\sim 10^{44}$ erg/s),
and the absence of optical signs of Seyfert activity.
Tidal disruption of a star by a supermassive black hole is the
favored explanation of these unusual events.
The second part provides a review of the 
initial X-ray observations, follow-up studies,
and the relevant aspects of tidal disruption
models studied in the literature. 
}
\end{abstract}

\section{Introduction}

\subsection{The search for SMBHs at the centers of galaxies}

The study of supermassive black holes and their cosmological evolution
is of great interest for a broad range of astrophysical topics including 
facets of galaxy formation and  
general relativity.  
In the last few decades, a number of different methods were developed to search for  
supermassive black holes (SMBHs)  
in external galaxies.  
Their detection in large numbers would clarify our understanding of the early phases
of the evolution of galaxies. In {\em active} galactic nuclei SMBHs
are now generally believed to be the prime mover of the non-stellar activity.
X-ray observations in the near future are expected to  
offer the opportunity of detecting some of the distinctive features of
strong field gravity, thereby also providing the ultimate proof for the
existence of black holes in AGN. 

There is now strong evidence for the presence of massive dark objects
at the centers of many galaxies. Does this hold for {\em all} galaxies ?
If so, why are some SMBHs `dark' ?
Questions of particular interest in the context of galaxy/AGN evolution are:
When and how did the first SMBHs
form and how do they evolve ?
What fraction of galaxies have passed through an active phase,
and how many now have 
non-accreting and hence unseen
supermassive black holes at their centers
(e.g., Lynden-Bell 1969, Rees 1989)?

Several approaches were followed to study these questions.
Much effort has
concentrated on the determination of central object masses from measurements
of the {\sl dynamics of
stars and gas} in the nuclei of nearby galaxies.
Earlier (ground-based) evidence for central quiescent dark masses
in galaxies 
(Kormendy \& Richstone 1995)    
has been strengthened
by recent HST results
(see Kormendy \& Gebhardt 2001 for a review).

A quite accurate determination of black hole mass was enabled by
the detection of water vapor maser emission
from the mildly active galaxy NGC\,4258 (Miyoshi et al. 1995). 
The water masers, whose motion can be precisely mapped
with VLBI, are located in a very compact
disk in Keplerian rotation around the central SMBH.
The fortunate geometry of the disk, nearly edge-on,  
allows to obtain the BH mass with high accuracy:
$M_{\rm BH} = 3.6\,10^{7}$ M$_{\odot}$ (Neufeld \& Maloney 1995, Greenhill et al. 1995).  

Still closer to the SMBH, in {\em active} galaxies with
with broad line region (BLR hereafter), the technique of
BLR reverberation mapping (e.g., Peterson 2001) provides
a powerful tool to estimate the BH mass via the clouds'
distance from the center and their velocity field
(e.g., Peterson \& Wandel 2000, Ferrarese et al. 2001).

\subsection{X-ray probes of the black hole region of AGN}

Whereas the dynamics of stars and gas probe rather large
distances from the SMBH,
high-energy {\sl X-ray emission}
originates from the immediate vicinity of
the black hole.
In {\em active} galaxies, excellent evidence for the presence of SMBHs
is provided by the detection of luminous hard power-law like X-ray emission,
rapid variability, and the discovery of evidence for  
relativistic effects in the iron-K line profile.
X-ray observations currently provide the most powerful
way to explore the black hole region of
AGN. 

X-rays at the centers of AGN arise in the
accretion-disk -- corona system (e.g, Mushotzky et al. 1993, Svensson et al. 1994,
Collin et al. 2000,
and references therein).
On larger scales, but still within the central region, X-rays might be emitted
by a hot intercloud medium at distances of the broad or narrow-line region
(e.g., Elvis et al. 1990).

The X-rays which originate from the accretion-disk region
are reprocessed in form of absorption and partial re-emission
(e.g., George \& Fabian 1991, Netzer 1993, Krolik \& Kriss 1995, 
Collin-Souffrin et al. 1996, Komossa \& Fink 1997b)
as they
make their way out of the nucleus.
The reprocessing bears the disadvantage of veiling the {\em intrinsic}
X-ray spectral shape,
and the spectral disentanglement of many different potentially contributing
components is not always easy.
However, reprocessing also offers the unique chance
to study the physical conditions
and dynamical states of the reprocessing material (see Komossa 2001 for a review), like: 
the outer parts of the accretion disk; the ionized absorber;
the torus, which plays an important role in AGN unification schemes
(Antonucci 1993); and the BLR and NLR.
Detailed modeling of the reprocessor(s) is also necessary
to recover the shape and properties of the {\em intrinsic} X-ray spectrum.

\begin{figure}[ht]
\caption{{Sketch of the central region of Seyfert galaxies. 
The black hole and
accretion disk region is surrounded by two systems of gas clouds, the broad line
region (BLR) and narrow-line region (NLR). These show up by their characteristic line emission
in optical spectra of AGN, and their presence is usually used to identify and classify
AGN. The molecular torus, and variants of it, are thought to play
an important role in unification schemes of Seyfert galaxies by blocking the
direct view on the BLR for certain viewing directions of
the observer (Antonucci 1993). Somewhere outside the bulk of the BLR, a relatively recently discovered
component of the active nucleus is located, the so-called
`warm' or ionized absorber (WA). \newline
A modification of this picture was recently proposed by Elvis (2000, 2001).
In his model, the BLR clouds arise from a flow of gas which rises vertically from
a narrow range of radii from the accretion disk. The flow then bends
and forms a conical wind moving radially outwards. The BLR clouds are
identified with the cool phase of this two-phase medium. Warm absorbers
appear if we view the continuum source through the wind (Elvis 2000, see his Fig.\,1)}
}
 \end{figure}

Recent progress has been made based on the improved spectral
resolution of the new generation of X-ray observatories, {\sl Chandra}
and {\sl XMM-Newton}. Both missions have imaging detectors and
grating spectrometers aboard. Their energy sensitivity bandpass covers
$\sim$(0.1--10) keV.
Below, a short review of results from these observatories is given,
starting at relatively large distances from the SMBH (NLR),
and then moving further inward (warm absorber and accretion disk region).

\paragraph{X-ray emission lines, X-ray narrow-line region.}
The detection of a high-temperature,
narrow-line, X-ray emitting plasma in NGC\,4151
was reported by Ogle et al. (2000),
confirming earlier evidence for extended X-ray emission from this galaxy
(Elvis et al. 1983).
 The X-ray gas is spatially coincident
with the NLR and extended narrow-line region.
For the first time, numerous emission-lines were detected in the
X-ray spectrum of NGC\,4151 with the HETG (High Energy Transmission
Grating Spectrometer) aboard {\sl Chandra}. 

\begin{figure}[t]
\hspace*{0.3cm}
\psfig{file=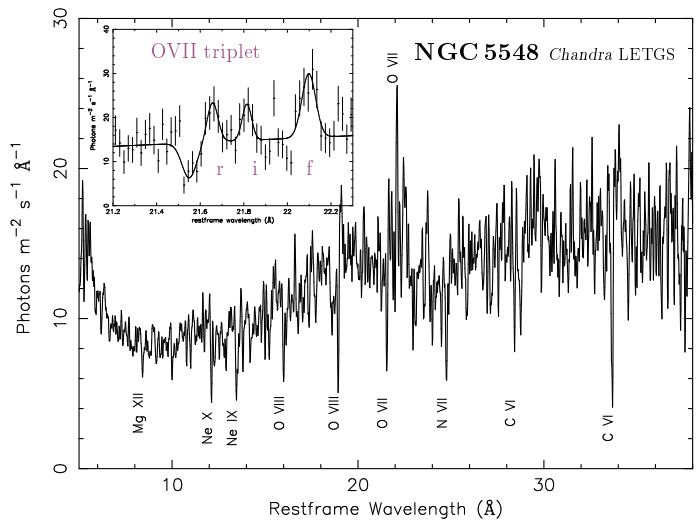,width=9.5cm,clip=}
\caption{{\sl Chandra} LETGS X-ray spectrum of NGC\,5548 (Kaastra et al. 2000).
The inset shows a zoom of the OVII triplet to which
a resonance line, two intercombination lines (unresolved), and a dipole-forbidden line
contribute.}
\label{wa}
 \end{figure}

The X-ray emission lines detected in the spectrum of NGC\,4151
and several Seyfert\,2 galaxies (e.g., Mrk\,3, NGC\,1068),
contain important
information on the physical conditions in the line-emitting medium,
like temperature, density,
and the main gas excitation/ionization mechanism - photoionization
or collisional ionization.       
Of particular importance in determining the 
main power mechanism of the lines,  
are the Helium-like triplets (Gabriel \& Jordan 1969; see our Fig.\ref{wa}),
the widths of the radiative recombination continua, and the
strengths of the Fe-L complexes (e.g., Liedahl et al. 1990).  

\paragraph{X-ray absorption lines, ionized absorber.}
With {\sl ROSAT}, the signatures of so-called `warm' absorbers, absorption edges
of highly ionized oxygen ions at
$E_{\rm OVII}=0.74$ keV and $E_{\rm OVIII}=0.87$ keV, 
were first detected in MCG$-$6-30-15
(Nandra \& Pounds 1992), following earlier {\sl Einstein} evidence for
highly ionized absorbing material in AGN 
(Halpern 1984). Detailed studies of many
other AGN followed, and the signatures of warm absorbers have now
been seen in about 50\% of the well-studied
Seyfert galaxies (see Komossa 1999 for a review).
First constraints place the bulk of the ionized material outside
the BLR, and depending on its covering factor and location, the warm absorber
may be one of the most massive components of the active nucleus. 
Evidence for ionized absorption was also found
in some very high-redshift quasars, starting with 
observations of  
PKS\,2351-154 (Schartel et al. 1997).
Some (but not all) warm absorbers were suggested to contain dust,
based on otherwise contradictory optical--X-ray observations
(e.g., Brandt et al. 1996, Komossa \& Fink 1997b, Komossa \& Bade 1998). 
The first possible detection of Fe-L dust features in the X-ray spectra
of MCG$-$6-30-15 and Mrk\,766 was recently reported by
Lee et al. (2001) and Lee (2001).   

The high-resolution spectrum of the Seyfert galaxy NGC\,5548, 
obtained with the {\sl Chandra} Low Energy Transmission Grating Spectrometer (LETGS),
shows many narrow absorption lines
of highly ionized metal ions of ogygen, neon, iron, etc. (Fig. \ref{wa}), 
confirming the presence of
a warm absorber in this galaxy (Kaastra et al. 2000). 
Similar signatures of ionized material have 
been detected with {\sl Chandra} and {\sl XMM-Newton}
in several AGN, including NGC\,3783 (Kaspi et al. 2000),
IRAS 13349+2438 (Sako et al. 2001),
NGC\,4051 (Collinge et al. 2001), Mrk\,509 (Yaqoob et al. 2002),
and MCG\,$-$6-30-15 (see next paragraph). 
First results show that the ionized absorption is complex with 
a range in ionization states.   

Assuming that the ionized absorber outflow is driven
by radiation pressure of the central continuum sources,
Morales \& Fabian (2001) demonstrated that observations
can then be used for an estimate of black holes masses
in AGN. They derive masses of $M_{\rm BH} \simeq 10^{6.5-7}$ M$_{\odot}$
for the galaxies of their sample.  

\begin{figure}[t]
\hspace*{0.6cm}
\psfig{file=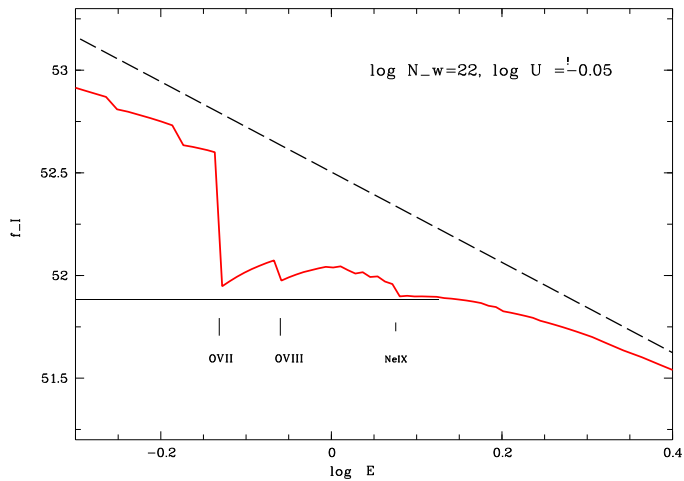,width=9.0cm,clip=}
\caption{Soft X-ray spectrum of an AGN, plotted as log flux [a.u.]
versus log Energy [keV]. The dashed line shows the input continuum spectrum. 
The thick solid line gives the spectrum after passage of a warm absorber. 
The calculation was carried out with the photoionization code {\sl Cloudy}
(Ferland 1993). Input parameters (ionization parameter $U$, column density $N_{\rm w}$)
were chosen similar to those obtained from a Beppo-SAX observation of Mrk\,766
(Matt et al. 2000)
except that $U$ was lowered. Only the absorption edges are shown, and are labeled in the
graph (absorption lines were omitted).  
If this theoretical absorption spectrum is now re-fit, without knowledge of the
intrinsic continuum (its shape and level), two fundamentally different solutions
are possible: (i) a high-level steep continuum in which case the absorption-solution
is recovered, or (ii) a low-level flat continuum (the horizontal thin solid line
in the graph) in which case the presence of a strong soft excess (at lower energies than
the OVII edge) plus some very broad emission lines are inferred.     
}
\label{m766}
 \end{figure}

\paragraph{Accretion-disk region, Fe-K line.}

The most direct probe of the black hole region, 
and particularly, of special and general relativistic effects,  
is emission from the inner part of the accretion disk (see Fabian 2001 for a review).
Tanaka et al. (1995) reported the detection of a
broadened FeK$\alpha$ line in MCG$-$6-30-15. The line profile
is well explained by the special relativistic effects of beaming and transverse
Doppler effect, and the general relativistic effect of gravitational redshift.      
Depending on details of modeling
the continuum, broad-winged Fe lines  may also be present 
in several XMM spectra of AGN (Nandra 2001). At certain times, the red wing
of MCG$-$6-30-15 is very broad, extending down to very soft energies (Wilms et al. 2001).

With {\sl XMM} it has also become clearer
that the Fe-line profiles are complex, and the line has several
sides of formation, including likely the BLR (NGC\,5548), the
torus (NGC\,3783, Mrk\,205), the X-ray ionization cone of NGC\,1068,
and a contribution from the outer parts of the accretion disk (MCG$-$6-30-15).

\paragraph{The case of MCG\,$-$6-30-15, X-ray spectral complexity.}
Whereas the signature of an ionized absorber in the form of narrow
absorption lines was detected in this galaxy with both,
XMM-Newton (Branduardi-Raymont et al. 2001) and {\sl Chandra}
(Lee et al. 2001),
new interpretations of some of the spectral features were put forward:
 Branduardi-Raymont et al. suggested
that the dominant soft X-ray features, so far interpreted
as metal absorption edges of the warm absorber, can be better
understood in terms of relativistically broadened  emission lines which
originate in the accretion disk.
On the other hand,
Matsumoto \& Inoue (2001) noted that the {\sl ASCA}-detected
broad wing of the iron K line - so far
interpreted in terms of relativistic broadening due to the
line's origin in the inner parts of the accretion disk -
could be successfully modeled by invoking a two-component warm absorber.

Fig. \ref{m766}
summarizes and visualizes one of the basic underlying ideas in
the discussion of emission versus absorption features
at soft X-ray energies (for many additional details
see Branduardi-Raymont et al. 2001 and Lee et al. 2001): Depending on where
the continuum is placed in Fig. \ref{m766}, one would either
infer the presence of {\sl a huge soft excess plus broad emission lines}, or a powerlaw
spectrum modified by {\sl absorption edges}.
We do not discuss these ideas further here, except for
noting that a disk-line interpretation of the bulk
of the soft X-ray features of MCG$-$6-30-15 would leave the puzzle
of the discrepant optical and X-ray absorption of this galaxy unanswered, which
could be solved by introducing a dusty warm absorber
(Reynolds et al. 1997).

Deep high-resolution X-ray observations of,
and a search for, variability of the spectral features will be
a very important next step in disentangling all components 
of the complex X-ray spectrum of this galaxy.

\subsection{The X-ray search for SMBHs in ULIRGs and \newline
            LINERs}

It is now generally believed that {\em active} galactic nuclei (AGN) are powered
by accretion onto SMBHs.
The search for heavily obscured SMBHs in ultraluminous infrared galaxies (ULIRGs),
and for low-luminosity AGN (LLAGN) in LINER galaxies is another interesting topic.
It will only be briefly touched here, since 
the emphasis of this review will be on recent evidence for SMBHs
in non-active `normal' galaxies (next Section).

\subsubsection{ULIRGs}

ULIRGs, characterized by their huge power-output in the infrared
which exceeds 10$^{12}\,L_\odot$ (Sanders \& Mirabel 1996),
are powered by massive starbursts or SMBHs. The discussion,
which one actually dominates received a lot
of attention in recent years (e.g., Joseph 1999, Sanders 1999).
In particular, only a small fraction of ULIRGs show AGN signatures
in their optical and infrared spectra.  Do the remaining ones 
nevertheless harbor AGN ? X-ray variability and luminous 
hard X-ray emission are excellent indicators of obscured AGN activity.  

With a redshift $z=0.024$ and a far-infrared
luminosity of $\sim 10^{12} L_{\odot}$,
NGC\,6240 is one of the nearest members of the class of ULIRGs.
Whereas X-rays from distant Hyperluminous IR galaxies, HyLIRGs,
were not detected by Wilman et al. (1999),
and the ULIRGs in the study of Rigopoulou et al. (1996)
were X-ray weak, NGC\,6240 turned out to be exceptionally
X-ray luminous. 
Starburst-driven superwinds are the most likely interpretation
of the extended emission (see Schulz \& Komossa 1999 for alternatives),
albeit being pushed to their limits to explain the huge
power output (Schulz et al. 1998).
The hard spectral component present in the {\sl ROSAT}
energy band was interpreted as scattered emission from an obscured 
AGN (Schulz et al. 1998, Komossa et al. 1998)
which shows up more clearly at higher energies, up 
to 100\,keV (e.g., Vignati et al. 1999, Mitsuda 1995, 
Ikebe et al. 2000). The intrinsically luminous AGN ($L_{\rm x} \approx 10^{44}$ erg/s),
can account for at least a substantial
fraction of the FIR power output of NGC\,6240.  

Using {\sl ASCA}, Nakagawa et al. (1999) studied the hard X-ray properties of 
a sample of 10 ULIRGs. Among these, 50\% have hard X-ray detections. 
The most stringent upper limit for the presence of any hard X-ray
emission was reported for Arp\,220. The possibility that it is 
a Compton-thick source cannot be excluded, though.

\subsubsection{LINERs}
LINER (Low-Ionization Nuclear Emission-Line Region)  galaxies
are characterized by their optical emission line spectrum
which shows a lower degree of ionization
than AGN. 
Their major power source and line excitation mechanism
have been a subject of lively debate ever since their discovery.
LINERs manifest the most common type of activity in the local universe.
If powered by accretion, they probably represent the low-luminosity end
of the quasar phenomenon,
and their presence has relevance to, e.g., the evolution of quasars,
the faint end of the Seyfert luminosity function, the soft X-ray background,
and the presence of SMBHs in nearby galaxies.

The X-ray properties of LINERs are inhomogeneous.
Spectra of a sample of objects studied by Komossa et al. (1999)
are best described by a composition
of soft thermal emission
and a powerlaw with varying relative
contributions of the two components from object to object.
Several studies of individual objects are consistent with these results
(e.g., Mushotzky 1982, Koratkar et al. 1995, Cui et al. 1997, 
Ptak et al. 1999, Roberts et al. 1999). 
X-ray luminosities are in the range $\sim$10$^{38-41}$ erg/s;
below those typically observed for Seyfert galaxies. 
The general absence of short-time scale (hours-weeks) X-ray variability
(Ptak et al. 1998, Komossa et al. 1999) is consistent with the suggestion that LINERs accrete in the
advection-dominated mode (e.g, Yi \& Boughn 1998, 1999, and references therein).
However, clear positive X-ray detections of LLAGNs in LINERs are still
rare. One potential problem problem is to distinguish 
powerlaw emission of the X-ray binary population of the host galaxy
from that of a genuine LLAGN. 
First {\sl Chandra} results on LINERs show that few, if any, are
obscured by absorbers of high column density (Ho et al. 2001).
Four out of eight LINERs of that study possess compact nuclear cores, consistent 
with AGNs.

\subsection{The X-ray search for SMBHs in {\itshape non-active} (`normal') galaxies, and tidal disruption
flares as probes } 

How can we find {\em dormant} SMBHs in {\em non-active} galaxies ?
Lidskii \& Ozernoi (1979) and Rees (1988)
suggested to use the flare of electromagnetic radiation produced
when a star is tidally disrupted and accreted by a SMBH
as a means to detect SMBHs in nearby {\em non-active} galaxies.

A star on a near-radial `loss-cone' orbit gets tidally disrupted 
once the tidal gravitational forces exerted by the black hole 
exceed the self-gravitational force 
of the star
(e.g., Hills 1975, Lidskii \& Ozernoi 1979,
Diener et al. 1997).
The tidal radius is given by
\begin{equation}
r_{\rm t} \simeq 7\,10^{12}\,({M_{\rm BH}\over {10^{6} M_\odot}})^{1 \over 3} 
    ({M_{\rm *}\over M_\odot})^{-{1 \over 3}} {r_* \over r_\odot}~{\rm cm}\,.
\end{equation}
The star is first heavily deformed, then disrupted.
About 50\%--90\% of the gaseous debris becomes unbound and is
lost from the system (e.g., Young et al. 1977, Ayal et al. 2000).
The rest will eventually be accreted by the black hole
(e.g., Cannizzo et al. 1990, Loeb \& Ulmer 1997).
The stellar material, first spread over a number of orbits,
quickly circularizes (e.g., Rees 1988, Cannizzo et al. 1990)
due to the action of strong
shocks when the most tightly bound debris interacts with
other parts of the stream (e.g., Kim et al. 1999).
Most orbital periods will then be within a few times
the period of the most tightly bound matter
(e.g., Evans \& Kochanek 1989; see also Nolthenius \& Katz 1982, Luminet \& Marck
1985).

Explicit predictions of the emitted spectrum and luminosity
during the disruption process and the start of the accretion
phase are still rare (see Sect.\,3 for details).
The emission is likely peaked in the soft X-ray or UV portion
of the spectrum, initially (e.g., Rees 1988,
Kim et al. 1999, Cannizzo et al. 1990; see also Sembay \& West 1993).

\begin{table*}[ht]
\caption{Summary of the X-ray and optical properties of the flaring normal
galaxies during outburst.
$z$ gives the redshift,
 $T_{\rm bb}$ is the black body
temperature derived from a black body fit to the X-ray high-state spectrum 
(cold absorption was fixed to the Galactic value in the direction
of the individual galaxies), `no emi.' means: no optical
emission lines were detected. $L_{\rm x,bb}$ gives the intrinsic luminosity in the
(0.1--2.4) keV band, based on the black body fit.
This is a lower limit to the actual peak luminosity,
since we most likely have not caught the sources exactly at maximum
light, since the spectrum may extend into the EUV, and since
it was conservatively assumed that no additional X-ray absorption occurs
intrinsic to
the galaxies.  
}
\vskip0.1cm
\begin{center}
\begin{tabular}{ccccc}
  \noalign{\smallskip}
  \hline
  \noalign{\smallskip}
galaxy name & $z$ & opt. type & $kT_{\rm bb}$ [keV] & $L_{\rm x,bb}$ [erg/s]  \\
  \noalign{\smallskip}
  \hline
  \hline
  \noalign{\smallskip}
NGC\,5905 & 0.011 & HII & 0.06 & 3 10$^{42}$$^*$  \\
  \noalign{\smallskip}
          &       &                 &                 \\ 
RXJ1242$-$1119  & 0.050 & no emi. & 0.06 & 9 10$^{43}$~  \\
         &       &      &                  \\
  \noalign{\smallskip}
RXJ1624+7554 &  0.064 & no emi. & 0.097 & $\sim$ 10$^{44}$~  \\
         &       &      &                  \\
  \noalign{\smallskip}
RXJ1420+5334 &  0.147 & no emi. & 0.04 & 8 10$^{43}$~ \\
          &       &      &                  \\
  \noalign{\smallskip}
RXJ1331$-$3243  & 0.051 & no emi. &  &  \\
  \noalign{\smallskip}
  \noalign{\smallskip}
\hline
\end{tabular}
\end{center}
  \noindent{\scriptsize $^{*}$Mean luminosity during the outburst; since the flux
 varied by a factor $\sim$3 during the observation, the peak luminosity is
higher. }
\end{table*}

\section{\bf Giant-amplitude X-ray flares from {\itshape non-active} galaxies}

\subsection{Summary of the original X-ray observations}

With the X-ray satellite {\sl ROSAT},
some rather unusual observations have been made in the last few
years: the detection of giant-amplitude, non-recurrent X-ray
outbursts from a handful of {\em optically non-active} galaxies,
starting with the case of NGC\,5905 (Bade et al. 1996, Komossa \& Bade 1999).
Based on the huge observed outburst luminosity,
the observations were interpreted in terms of tidal disruption events.
Below, we first give a brief review of the properties of all published
X-ray flaring non-active galaxies
and then discuss the favored outburst scenario.
A Hubble constant of $H_0 = 50$ km/s/Mpc is adopted throughout the paper.

\begin{figure}[t]
\psfig{file=komossa_f4a.ps,width=5.8cm,clip=}
\vspace*{-5.8cm}\hspace{6.1cm}
\psfig{file=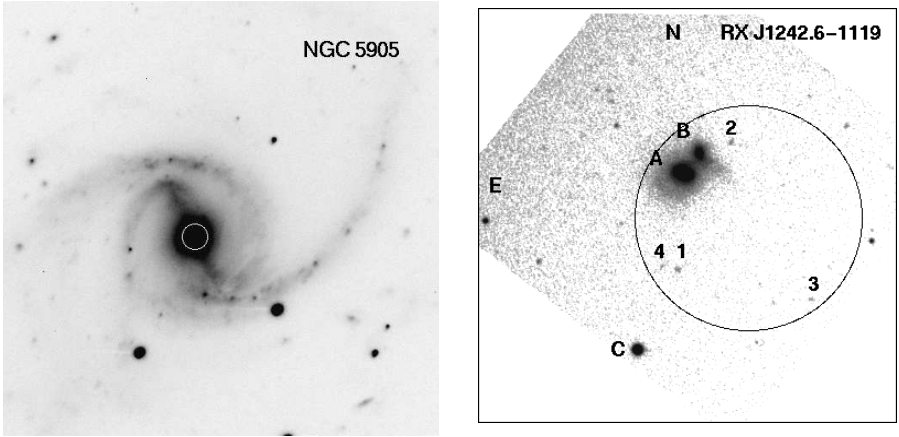,width=6cm,clip=}
\caption{Optical images of NGC\,5905 (right) and RXJ1242$-$1119 (left).
The circles mark the positional uncertainty of the X-ray emission.}  
\label{opt_ima}
 \end{figure}

So far, four X-ray flaring {\em non-active} galaxies have been detected:
NGC\,5905 (Bade et al. 1996, Komossa \& Bade 1999; see Fig. \ref{opt_ima}),
RXJ1242$-$1119 (Komossa \& Greiner 1999), 
RXJ1624+7554 (Grupe et al. 1999),
and RXJ1420+5334 (Greiner et al. 2000){\footnote{The
X-ray position error circle of RXJ1420+53 contains a second galaxy 
for which an optical spectrum
is not yet available. Based on the galaxy's morphology, Greiner et al. argue
that it is likely non-active.}};
first results on a fifth 
candidate were presented by Reiprich \& Greiner (2001). Based on the position they report,
we refer to this source as RXJ1331$-$3243. 
All these galaxies show similar
properties:
\begin{itemize}

\item
huge X-ray peak luminosity (up to $\sim 10^{44}$ erg/s),

\item
giant amplitude of variability (up to a factor $\sim$ 200),

\item
ultra-soft X-ray spectrum ($kT_{\rm bb} \simeq$ 0.04--0.1 keV
when a black body model is applied).

\end{itemize}

\noindent A summary of the observations is provided in Table 1.
In Fig. \ref{light} X-ray lightcurves
of NGC\,5905 and RXJ1420+53 are overplotted, shifted in time to the
same date of outburst to allow direct comparison.
So far, the best sampled lightcurve is that of NGC\,5905.
The `merged' lightcurve is consistent with a fast rise
and a decline on a time scale of months to years.
We performed a preliminary analysis of an archival {\sl ASCA} observation
of NGC\,5905 carried out in 1999. The flux of NGC\,5905 did not
drop further compared to the last {\sl ROSAT} observations.

\subsection{Optical observations}

Given the unusual X-ray properties of these galaxies, a very
important question was: what is the optical classification
of the flaring galaxies. Particularly: are there any hints of weak
permanent Seyfert activity ? This question is of great
interest when discussing outburst scenarios{\footnote{For instance,
theorists working on tidal disruption of stars repeatedly argued 
that it would be important to be certain about
the (optically) {\em non-active} nature of the flaring galaxy,
to exclude AGN-related variability mechanisms (changes in the accretion disk).}}.

Optical spectra were taken at different times 
and at different telescopes, starting several years after the X-ray high-states. 
The optical spectrum of NGC\,5905 turned out to be of HII-type,
consistent with its classification prior to the observed X-ray outburst.
Neither broad wings in the Balmer lines, nor AGN-typical emission-line ratios,
nor high-ionization lines, which usually indicate the presence of an AGN,
were detected. 
The spectra of the other galaxies 
only show absorption lines from the host galaxies.

\begin{table*}[h]
\caption{Coordinates (J\,2000) of the optical centers of the galaxies
identified as counterparts to the X-ray flares (NGC\,5905, RXJ1242$-$1119A, RXJ1624+7554)
or X-ray position (RXJ1420+5334, RXJ1331$-$3243).} 
\vskip0.1cm
\begin{center}
\begin{tabular}{cl}
  \noalign{\smallskip}
  \hline
  \noalign{\smallskip}
galaxy name & coordinates  \\
  \noalign{\smallskip}
            & RA ~~~~~~~~~~~~~~~DEC   \\
  \noalign{\smallskip}
  \hline
  \hline
  \noalign{\smallskip}
NGC\,5905 & 15$^{\rm h}$15$^{\rm m}$23.4$^{\rm s}$~~~~+55$^{\rm o}$31$\arcmin$02$\arcsec$  \\
  \noalign{\smallskip}
RXJ1242$-$1119A & 12$^{\rm h}$42$^{\rm m}$38.5$^{\rm s}$~~~~$-$11$^{\rm o}$19$\arcmin$21$\arcsec$  \\
  \noalign{\smallskip}
RXJ1624+7554 & 16$^{\rm h}$24$^{\rm m}$56.5$^{\rm s}$~~~~+75$^{\rm o}$54$\arcmin$56$\arcsec$  \\
  \noalign{\smallskip}
RXJ1420+5334 & 14$^{\rm h}$20$^{\rm m}$24.2$^{\rm s}$~~~~+53$^{\rm o}$34$\arcmin$11$\arcsec$  \\
  \noalign{\smallskip}
RXJ1331$-$3243 & 13$^{\rm h}$31$^{\rm m}$57.6$^{\rm s}$~~~~$-$32$^{\rm o}$43$\arcmin$20$\arcsec$  \\
  \noalign{\smallskip}
  \hline
\label{coo}
\end{tabular}
\end{center}
\end{table*}

\subsection{Radio observations}

Radio observations are important for two reasons:
Firstly, they allow the search for a peculiar, optically hidden AGN at the center of
each flaring galaxy
(as already discussed by Komossa \& Bade (1999)
this possibility is a very unlikely explanation for the X-ray
flares. It is very important, though, to safely {\em exclude}
exotic AGN scenarios).
Besides hard X-ray observations,
compact radio emission is a good indicator of AGN activity
because radio photons can penetrate even high-column density
dusty gas which is not transparent to optical or soft X-ray photons.
Secondly, radio emission could possibly be produced in relation to
the X-ray flare itself.

\subsubsection{NVSS and FIRST search for radio emission from the X-ray outbursters}

A search for radio emission from the X-ray flaring
galaxies was performed, utilizing 
the NRAO VLA Sky Survey (NVSS) catalogue (Condon et al. 1998) which contains
the results of a 1.4\,GHz radio sky survey north of $\delta$=--40$^{\rm o}$.
Except NGC\,5905, no flaring galaxy has a NVSS detection.
At 1.4\,GHz, the emission of NGC\,5905 appears extended and is thus related
to the galaxy instead of the nucleus (see also next Section).

NGC\,5905 is also detected in the FIRST VLA sky survey at 20cm
(e.g., Becker et al. 1995).
No radio emission from RXJ1420+53 was found.
The FIRST catalogue detection
limit at the source position is 0.96 mJy/beam.
None of the other outbursters is located
within a FIRST survey field.

\subsubsection{Radio emission from NGC\,5905}

21 cm neutral hydrogen line emission was detected by van Moorsel (1982; see also
Staveley-Smith \& Davies 1987),
using the Westerbork Synthesis Radio Telescope (WSRT).
The emission is spatially resolved (Fig. 3 of van Moorsel 1982) with an extent of diameter
7.3\arcmin. Peaks in the HI emission closely follow the spiral arms.
Whereas the bulk of the radio emission detected at the frequency
of the 21cm line is unrelated to the nucleus,
 van Moorsel also briefly mentions the presence of unresolved
continuum emission of 13.2 mJy.

\begin{figure*}
\hspace*{-0.8cm}
\psfig{file=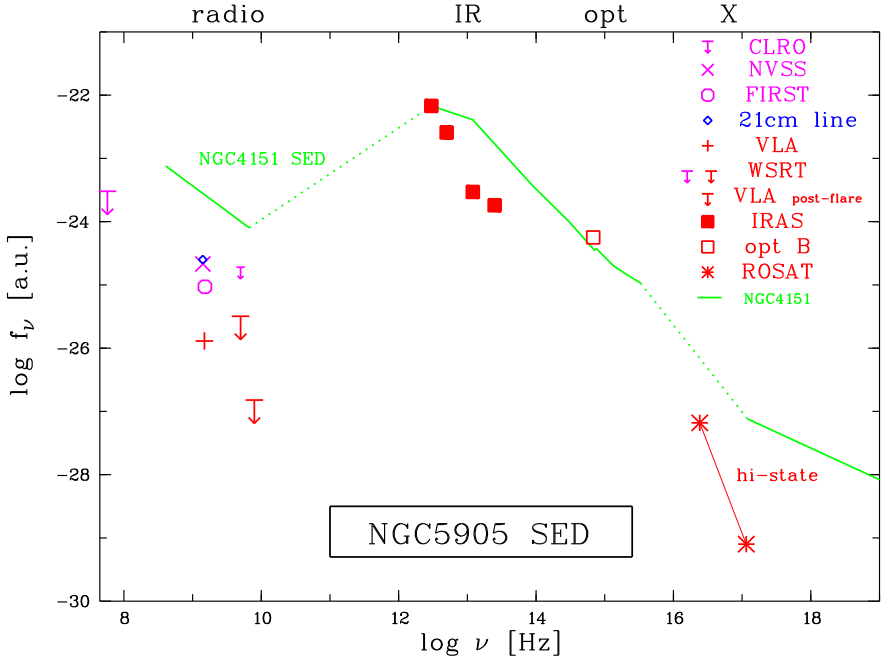,width=13.0cm,angle=270}
\caption[sed]{Multi-wavelength continuum spectrum (SED) of NGC\,5905
(symbols). 
The SED of the active
galaxy NGC\, 4151 (Komossa 2001) is shown for comparison by the solid/dotted line.
Arrows denote upper limits.
Radio data, from left to right:
arrow: Clark Lake Radio Observatory TPT array (Israel \& Mahoney 1990);
lozenge: 21 cm line(van Moorsel 1982);
cross: NVSS survey; open circle: FIRST survey;
plus: VLA (Hummel et al. 1987);
upper limits: WSRT (upper point: total emission within 2.4\arcmin,
lower point: nuclear emission within 10\arcsec; Brosch \& Krumm 1984);
upper limit: VLA post-flare observation (Komossa \& Dahlem 2001);
filled squares: IRAS (taken from NED data base), open square: optical B-magnitude
(taken from NED),
asterisks: X-ray high-state emission (Komossa \& Bade 1999).
   Note: data were taken with different
      aperture sizes and resolution, and at different times. 
}
\label{sed}
\end{figure*}

Extended radio emission was also found by Hummel et al. (1987) at 1.49 GHz,
whereas Brosch \& Krumm (1984) reported upper limits at 5 GHz
(for both, extended emission and a nuclear source). Israel \& Mahoney (1990)
give an upper limit at 57.5 MHz (see our Fig. \ref{sed}).

Hummel et al. (1987)
reported the presence of an unresolved ($\approxlt 2\arcsec$)
core source at a frequency of 1.49 GHz.
Similar sources were found in 41\% of the HII galaxies
of the `complete sample' of Hummel et al. (1987; their Tab.\,1).

Finally, we note that the NVSS and FIRST surveys
were performed after the X-ray outburst of NGC\,5905.
However, the NVSS value is consistent with previous measurements of
extended radio emission at that frequency. Similarly,
the FIRST value is consistent with the pre-flare measurement of
Hummel et al. (1987) at the same frequency.

In order to search for a radio source at the nucleus of NGC\,5905
after the X-ray outburst
(besides a permanent AGN, radio emission could be produced in relation
to the X-ray flare)
a radio observation with the VLA A array at 8.46 GHz
was carried out by M. Dahlem in 1996.
No radio source was detected within the central field
of view of 100$\arcsec$$\times$100$\arcsec$ with
a 5$\sigma$ upper limit
for the presence of a central point source of 0.15 mJy
(Komossa \& Dahlem 2001).
Assuming a distance of 75.4 Mpc of NGC\,5905 this
translates into an upper limit
of $L_{\rm 8.46GHz} \leq\ 1.0\,10^{20}$ W/Hz.

The radio measurements of NGC\,5905 are summarized in Fig. \ref{sed}
together with multi-wave\-length observations.

\subsubsection{Origin of the radio emission of NGC\,5905}

A large contribution to the radio emission of NGC\,5905 comes from
the 21cm line of neutral hydrogen.
In addition there is some radio continuum emission at
the same frequency, and at 1.49 GHz (FIRST).  No other
radio detection was reported.

The radio emission of large samples of spiral galaxies was examined by, e.g.,
Brosch \& Krumm (1984), Hummel et al. 1987,
Giuricin et al. (1990), Israel \& Mahoney (1990), Sadler et al. (1995),
and Falcke (2001).
Radio emission (extended and from the inner few arcseconds)
was generally detected from a number of the
non-active spiral galaxies in the samples.
E.g., Hummel et al. find unresolved ($\approxlt 2\arcsec$)
core sources in 41\% of the HII galaxies
of their `complete sample' at  1.49 GHz.
At 57.5 MHz, Israel \& Mahoney (1990) detected 68 out of 133 observed
galaxies.
Trends were repeatedly reported in the literature, that
paired HII galaxies (like NGC\,5905) and interacting galaxies
show enhanced total and central radio emission compared to isolated
galaxies (e.g., Giuricin et al. 1990,
and references therein).
Enhanced star-formation activity was considered a possible
explanation of this effect.
In some cases, nuclear radio sources in spirals could possibly be
explained by radio supernovae (Sadler et al. 1995).

In summary, the radio emission of NGC\,5905 is not unusual for its class 
and does not indicate the presence of a luminous, optically hidden  AGN{\footnote{The radio observations
do not exclude the presence of some low-level nuclear radio activity related to a
low-luminosity AGN, like it may be present for instance in many LINERs (e.g., Falcke 2001,
and references therein).}}.

\begin{figure}[ht]
\psfig{file=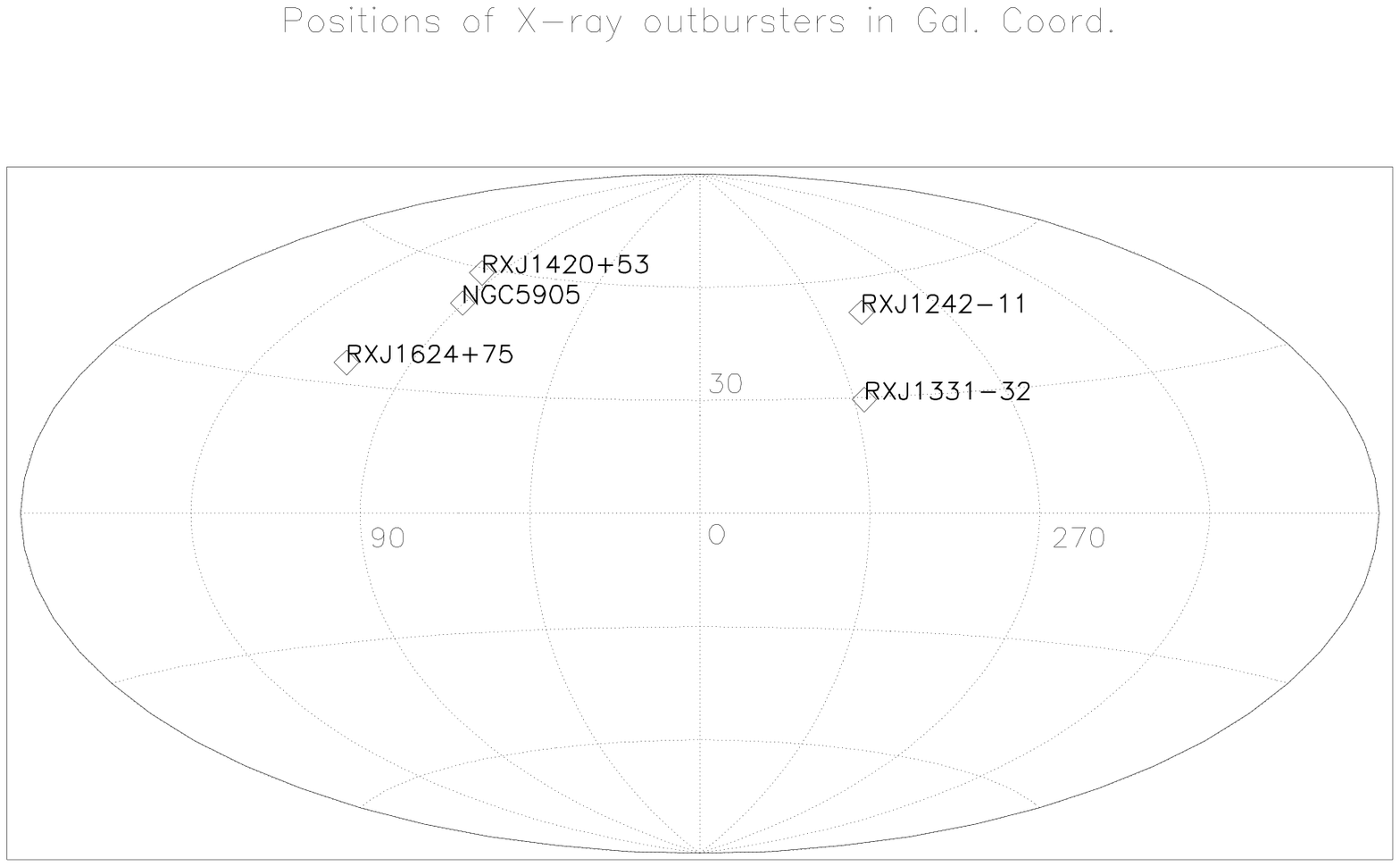,width=10.4cm,clip=}
\caption{Locations of the X-ray flaring galaxies on the sky (Galactic II coordinates
in aitoff projection).
}
\end{figure}

\section{Favored outburst scenario: tidal disruption of a star by a SMBH}

\subsection{(Rejected or unlikely) alternatives to tidal disruption: 
                 stellar sources, lensing, GRB, hidden AGN}

Most examined outburst scenarios do not survive
close scrutiny (Komossa \& Bade 1999), because they
cannot explain the huge maximum luminosity (e.g.,
X-ray binaries within the galaxies, or a supernova in a dense medium),
are inconsistent
with the optical observations (gravitational lensing),
or predict
a different temporal behavior (X-ray afterglow of a Gamma-ray burst;
see, e.g., Fig. 2 of Bradt et al. 2001).
Standard AGN scenarios cannot
account for the soft X-ray flares and the absence of optical
AGN-like emission lines (see the discussion by Komossa \& Bade 1999
and Komossa \& Voges 2001 for more details).

\begin{figure}[t]
\psfig{file=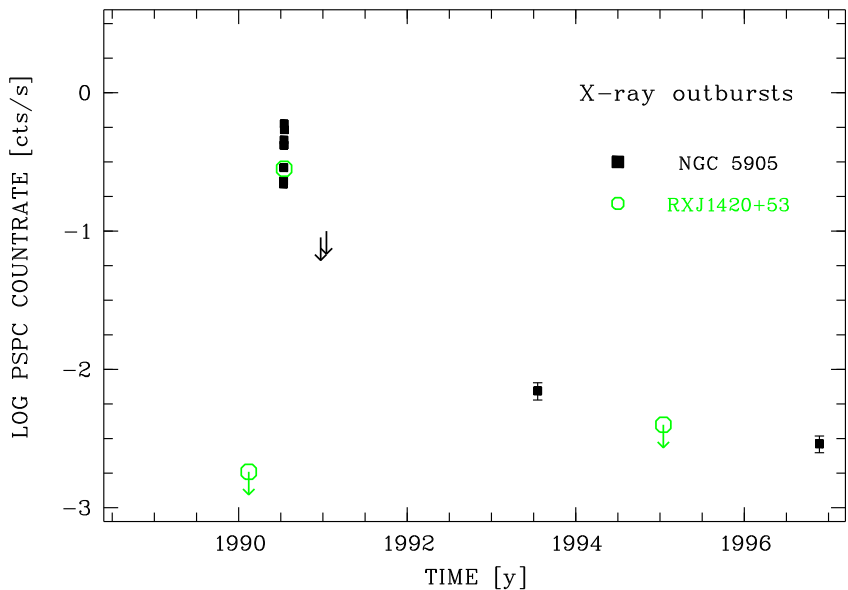,width=11.5cm,clip=}
\caption{X-ray light curve of NGC\,5905 (black squares), and RXJ1420+53
(open circles; shifted in time to match outburst date of NGC\,5905). Arrows
denote upper limits. A preliminary analysis of an archival {\sl ASCA} observation
of NGC\,5905 carried out in 1999 shows that the source flux did not
drop further compared to the last {\sl ROSAT} observations. }
\label{light} 
\end{figure}
%

\subsection{\bf Tidal disruption model}

Except possibly for some GRB-related emission mechanisms, the huge peak outburst
luminosity nearly inevitably calls for the presence of a SMBH{\footnote{in fact, even
most GRB scenarios involve the presence of a black hole as ultimate energy reservoir}}.
This, in combination with the complete absence of any signs of AGN activity
at all wavelengths,
makes tidal disruption of a star by a SMBH
the favored outburst mechanism.

After a short overview of aspects of tidal disruption models
discussed in the literature, we apply these models 
to the X-ray flare observations. 

\subsubsection{Tidal disruption, short overview}

Historically, tidal disruption of captured stars by black holes was first considered
in relation to star clusters and galactic nuclei (e.g., Frank \& Rees 1976), and was
applied to the nuclei of active galaxies where it was suggested as a means
of fueling AGN (e.g., Hills 1975, Sanders 1984), 
or to explain UV-X-ray variability
of AGN (e.g., Kato \& Hoshi 1978).
Shields \& Wheeler (1978)
then argued that 
tidal disruptions of captured stars cannot provide an effective source of fuel of AGN,
basically because of two problems:
Firstly, the disruption rate is not high enough
to sustain a permanent gas flow, and low-angular momentum orbits are quickly depleted of stars. 
If, however, the
stellar density
close to the black hole is high enough, stellar collisions would dominate
the gas supply over tidal disruption. Secondly, it is difficult to
account for the luminosity of the most luminous quasars,
since these have masses where the tidal radius is inside the Schwarzschild radius.

A star will only be disrupted if its tidal radius
lies outside the Schwarzschild radius of the black hole, else
it is swallowed as a whole. This happens for black hole masses larger than
$\simeq$10$^8$ M$_{\odot}$; in case of a Kerr black hole, tidal
disruption may occur for even larger BH masses if the star
approaches from a favorable direction (Beloborodov et al. 1992).
More massive black holes may still strip the atmospheres of giant stars.

Due to the complexity of the problem, theoretical work focussed
on different subtopics of the complete problem, and 
on stars of solar mass and
radius.
Calculations and numerical simulations of
 the disruption process,
the stream-stream collision, the accretion phase, 
the changes in the stellar distribution
of the surroundings and the depopulation and refilling of low-angular momentum orbits,
and the disruption rates have been studied in the
literature (e.g., Nolthenius \& Katz 1982, 1983,
Carter \& Luminet 1985, Luminet \& Marck 1985, Evans \& Kochanek 1989,
Laguna et al. 1993, Diener et al. 1997, Ayal et al. 2000, Ivanov \& Novikov 2001;
Kochanek 1994, Lee et al. 1995, Kim et al. 1999;
Hills 1975, Gurzadyan \& Ozernoi 1979, 1980, Cannizzo et al. 1990, Loeb \& Ulmer 1997,
Ulmer et al. 1998;
Frank \& Rees 1976, Lightman \& Shapiro 1977, Norman \& Silk 1983, Sanders \& van Oosterom 1984,
Rauch \& Ingalls 1998, Rauch 1999;
Syer \& Ulmer 1999, Magorrian \& Tremaine 1999).
DiStefano et al. (2001) recently considered the case of $M > M_{\odot}$, and suggested
that remnants of tidally stripped stars might be detected   
as supersoft X-ray sources at the centers of nearby galaxies. 

\begin{figure}[h]
\psfig{file=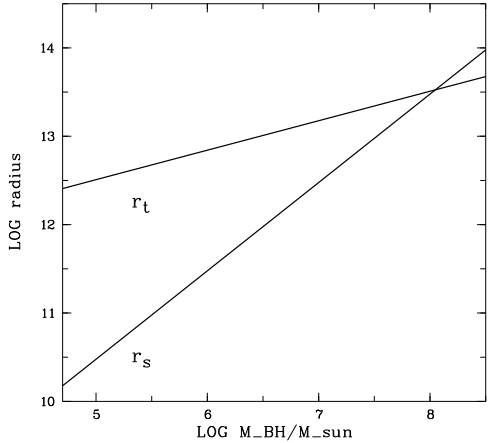,width=6.5cm,clip=}
\hfill
\begin{minipage}[]{0.35\hsize}\vspace*{-6.7cm}
\hfill
\caption{Run of tidal radius and Schwarzschild radius in dependence
of black hole mass, for a star of solar mass and radius. For large SMBH masses, stars are swallowed whole,
since the tidal radius is no longer outside the Schwarzschild radius.}
\end{minipage}
 \end{figure}

\subsubsection{Tidal disruption in {\itshape active} galaxies (?)} 

During the last several years,
tidal disruption was occasionally considered as explanation of  
some properties of {\em active} galaxies  
(either AGN as a class,
or individual peculiar observations), 
although alternative
interpretations existed in each case:
Tidal disruption was applied by Eracleous et al. (1995) in a duty cycle model
to explain the UV brightness/darkness of LINERs. Roos (e.g., 1992) suggested
an origin of the BLR clouds of AGN in terms of the gaseous debris of tidally
disrupted stars.
Peterson \& Ferland (1986) proposed tidal disruption as possible explanation for
the transient brightening and broadening of the HeII line observed in
the Seyfert galaxy NGC\,5548.
Brandt et al. (1995) and Grupe et al. (1995) reported the detection of an X-ray outburst
from the active galaxy Zwicky 159.034 (IC\,3599). Besides other outburst
mechanisms, tidal disruption was briefly mentioned as possibility.
Based on high-resolution post-outburst optical spectra,
Komossa \& Bade (1999) classified IC3599 as Seyfert type 1.9.
In the UV spectral region, two UV spikes were detected at and near
the center of the elliptical
galaxy NGC\,4552. The central flare, although rather weak,
 was interpreted by Renzini et al.
(1995) as accretion event (the tidal stripping
of a star's atmosphere by a SMBH, or the accretion of a molecular cloud).
There are several indications (e.g., from radio observations), that
NGC\,4552 shows permanent low-level activity. 

\subsubsection{Emission of radiation, temporal evolution, and model uncertainties}

Intense electromagnetic radiation will be emitted
in three phases of the disruption and accretion process:
First, during the stream-stream collision when different parts
of the bound stellar debris first interact with themselves (Rees 1988).
Kim et al. (1999) have carried out numerical simulations of this
process and find that the initial burst due to
the collision may reach a luminosity of 10$^{41}$ erg/s, under the assumption
of a BH mass of 10$^6$ M$_{\odot}$ and a star of solar mass and radius.
Secondly, radiation is emitted during the accretion of the stellar
gaseous debris. Finally, the unbound stellar material leaving the system
may shock the surrounding interstellar matter
and cause intense emission, like in a supernova remnant (Khokhlov \& Melia 1996).

Many details of the tidal disruption and the
related processes are still unclear.
In particular, the flares cannot be standardised. Observations
would depend on many parameters, like the type of disrupted star, the impact
parameter, the spin of the black hole, effects of relativistic precession,
 and the complication of radiative transfer 
by effects of viscosity and shocks (Rees 1990).
Uncertainties also include the amount of the stellar material 
that is accreted (part may be ejected as a thick wind, or
swallowed immediately). Related to this is the duration
of the flare-like activity, which may be months or years
to tens of years (e.g., Rees 1988, Cannizzo et al. 1990,
Gurzadyan \& Ozernoi 1979), followed by a decline on a longer time scale.
The flare duration depends on how fast the stellar debris circularizes
and how fast it accretes. If both time scales are short 
and the material accretes at the same rate as it falls back towards
the black hole, the decline time scale of the flare scales as
\begin{equation}
{dm \over{dt}} \propto t^{-{5\over{3}}}
\end{equation}
(e.g., Evans \& Kochanek 1989, Rees 1990). 
Else energy
output will be spread more evenly in time (Cannizzo et al. 1990). 

Another uncertainty in predicting the flare spectrum 
arises from the  mode of accretion:
does it proceed via a thin disk
(e.g., Cannizzo et al. 1990) or a thick disk (Ulmer et al. 1998),
and, if in a thick disk, under which angle do we view it ?

\subsection{Order of magnitude estimates and consistency checks}

\subsubsection{Inferences from X-ray observations}

Although many details of the actual tidal disruption process are still unclear,
some basic predictions have been repeatedly made in the literature
how a tidal disruption event should manifest itself observationally:
\begin{itemize}
\item
(1) the event should be of finite
duration (a `flare'), 
\item 
(2) it should be very luminous
(up to $L_{\rm max} \approx 10^{45}$ erg/s
in maximum), and 
\item (3) it should reside in a galaxy
which is otherwise perfectly {\em non}-active
(to be sure to exclude an upward fluctuation in
gaseous accretion rate of an {\em active} galaxy).
\end{itemize}
All three predictions are fulfilled by the X-ray flaring galaxies;
particularly by NGC\,5905 and RXJ1242$-$1119, which are the two best-studied
cases so far.

In addition, we can do some further order of magnitude estimates and consistency checks.
The luminosity emitted if the black hole is accreting at its Eddington luminosity
can be estimated by
\begin{equation}
L_{\rm edd} = {{4 \pi G M m_{\rm p} c}\over{\sigma_{\rm T}}} ~~\simeq 1.3 \times 10^{38} M/M_{\odot} ~{\rm erg/s}\,.
\end{equation}
In case of NGC 5905, a BH mass of at least a few $\sim$$10^{4}$ M$_{\odot}$ would be
required to
produce the observed luminosity.
This is a {\em lower limit} on the black hole mass, 
since we likely did not observe $L_{\rm x}$ at its peak
value due to observational gaps in the lightcurve, and since other
conservative assumptions were made{\footnote{e.g., the amount of absorption
was fixed to the lowest possible value, $N_{\rm Gal}$}}.
A comparison with the SMBH mass of NGC\,5905 using indirect {\em optical}
methods is performed in Sect. 3.3.2. 
For the other galaxies, using again $L_{\rm edd}$, we infer
BH masses reaching up to a few 10$^6$ M$_{\odot}$.
This is, again, a lower limit.
Alternative to a complete disruption event, the atmosphere of a giant star could have been
stripped.

In a simple black body approximation, the temperature
of the accretion disk scales with black hole mass
as
\begin{equation}
T \simeq 8\,10^4 \,({M_{\rm BH}\over {M_\odot}})^{1 \over 12} ~{\rm K} ~~ ({\rm at}~ r_t), ~~~T \simeq 2\,10^7 \,({M_{\rm BH}
\over {M_\odot}})^{-{1 \over 4}}~{\rm K} ~~ ({\rm at}~ 3\,r_S)\,.
\end{equation}
This gives $T_{\rm r_{tidal}} \simeq 3\,10^5$\,K, $T_{\rm 3r_S} \simeq 7\,10^5$\,K
for M=10$^6$\,M$_{\odot}$, where $r_{\rm S}$ is the  Schwarzschild radius.
Using black body fits of the X-ray flare spectra
we find
temperatures in a similar range; $T_{\rm obs} \simeq$ (4-10)\,10$^5$ K.
Like in AGN, X-ray powerlaw tails could develop. They might have
escaped detection during the observations, since weak,
or they may develop only after a certain time after the start
of the accretion phase.
We soon expect first results from a {\sl Chandra} and {\sl XMM} observation
of RXJ1242$-$1119, which will give valuable constraints on the post-flare evolution.

The Eddington time scale for the accretion of the stellar material is given by
\begin{equation}
t_{\rm edd} \simeq 4\,\eta_{0.1} (M_{\rm BH}/10^6M_{\odot}) (M_*/0.1M_{\odot}) ~{\rm yrs}\,.
\end{equation}
Uncertainties in estimating the total duration of the tidal disruption event
arise from questions like: how much material is actually accreted or expelled, does
a strong wind develop, etc. (see Sect. 3.2.3).
 The events are expected to last for months to years (e.g., Rees 1988).
Observationally, the duration of the events was at least several days, followed by gaps in the observations.
The source fluxes were then significantly down several years later (e.g., Fig.\,9 of Komossa \& Bade 1999).
Apart from theoretical uncertainties in the prediction of the decline time scale in 
the total luminosity output (Sect. 3.2.3),
the emission will also likely shift outside 
the {\sl ROSAT} band, from the X-ray to the EUV-UV
band, with increasing time.  

Finally, we note that the redshift distribution of the few sources observed
so far is consistent with the predicted tidal disruption rate,
in the sense that the events are sufficiently distant to define
a large volume of space, in which the detection of a few events would be expected.

\subsubsection{Non X-ray estimates of the black hole mass in NGC\,5905}

An optical rotation curve was obtained by Komossa \& Bade (1999)
which allowed an estimate of the mass enclosed within 0.7 kpc:
$M \approx 10^{10}$ M$_{\odot}$.
This immediately provides an upper limit on BH mass,
but
the volume sampled is still too large to estimate
the actual BH mass. An HST-based rotation curve  would significantly
improve the above limit and thus the constraints
on a central dark mass.

In order to get a better (non-X-ray) estimate of the
BH mass of NGC\,5905, we used
the correlation between bulge properties and BH mass.

NGC\,5905 has a total blue magnitude of m$_{B,0} = 12.1^{\rm m}$.
Using the bulge-to-disk luminosity ratio generally valid for galaxies
of the Hubble-type of NGC\,5905 (SBb), $k=0.25$ (Salucci et al. 2000),
gives the absolute bulge blue magnitude, B$_{\rm T,0}^{\rm bulge} = -20.5$.
We then compared with two recent studies that correlate BH mass
and bulge blue luminosity: (i) the work of Ferrarese \& Merritt (2000; FM00)
which concentrates mostly on elliptical galaxies,
and (ii) the work of Salucci et al. (2000; S00) on spiral galaxies.

Using the relation between B$_{\rm T,0}^{\rm bulge}$ and $M_{\rm BH}$
of FM00 (their `sample A', their Tab. 2; see also, e.g., 
Franceschini et al. 1998) gives a BH mass of
a few times 10$^{8}$ M$_{\odot}$ for NGC\,5905.
This is close to the limiting BH mass for tidal disruption of a solar-type
star to work.
However, it has to be kept in mind, that the B$_{\rm T,0}^{\rm bulge}$ - $M_{\rm BH}$
relation shows a very large scatter (in contrast to the $M_{\rm BH}$ - $\sigma$ relation),
and that the few spirals in the sample of FM01 tend to be located below the relation
followed by ellipticals.
Therefore, in a second step, we used the results of S00 on BH masses in
(late-type) spiral galaxies (their Fig. 6). In that case we obtain
an upper limit for the BH mass of NGC\,5905 of $M_{\rm BH} \approxlt 10^7$ M$_{\odot}$.

Finally, the relation between radio luminosity and BH mass of Franceschini et
al. (1998, their Fig.\,3; see also Wu \& Han 2000) was employed.
The measured radio upper limit for the nucleus of NGC\,5905 then translates   
into an upper limit on black hole mass of $M_{\rm BH} \approxlt 2.5\,10^8$ M$_{\odot}$. 
Results are summarized in Tab. \ref{mass-esti}.

\begin{table*}[ht]
\caption{Summary of mass estimates of the black hole at the center of NGC\,5905,
employing different methods (see the text for details).}
\vskip0.1cm
\begin{center}
\begin{tabular}{cll}
  \noalign{\smallskip}
  \hline
  \noalign{\smallskip}
energy band & method & BH mass  \\
  \noalign{\smallskip}
  \hline
  \hline
  \noalign{\smallskip}
X-rays & $L_{\rm Eddington}$ & $>$ few\,10$^4$ M$_{\odot}$ \\
  \noalign{\smallskip}
optical & $M_{\rm B,bulge} - M_{\rm BH}$  correlation, spirals &  $\approx 10^7$ M$_{\odot}$  \\
  \noalign{\smallskip}
        & $M_{\rm B,bulge} - M_{\rm BH}$  correlation, ellipticals & $\approx$ few\,10$^8$ M$_{\odot}$  \\ 
  \noalign{\smallskip}
        & rotation curve & $\ll 10^{10}$ M$_{\odot}$  \\
  \noalign{\smallskip}
radio   & nuclear radio power $-$ $M_{\rm BH}$  correlation & $< 2.5\,10^8$ M$_{\odot}$ \\
  \noalign{\smallskip}
  \hline
\label{mass-esti}
\end{tabular}
\end{center}
\end{table*}   

  \begin{figure}
\psfig{file=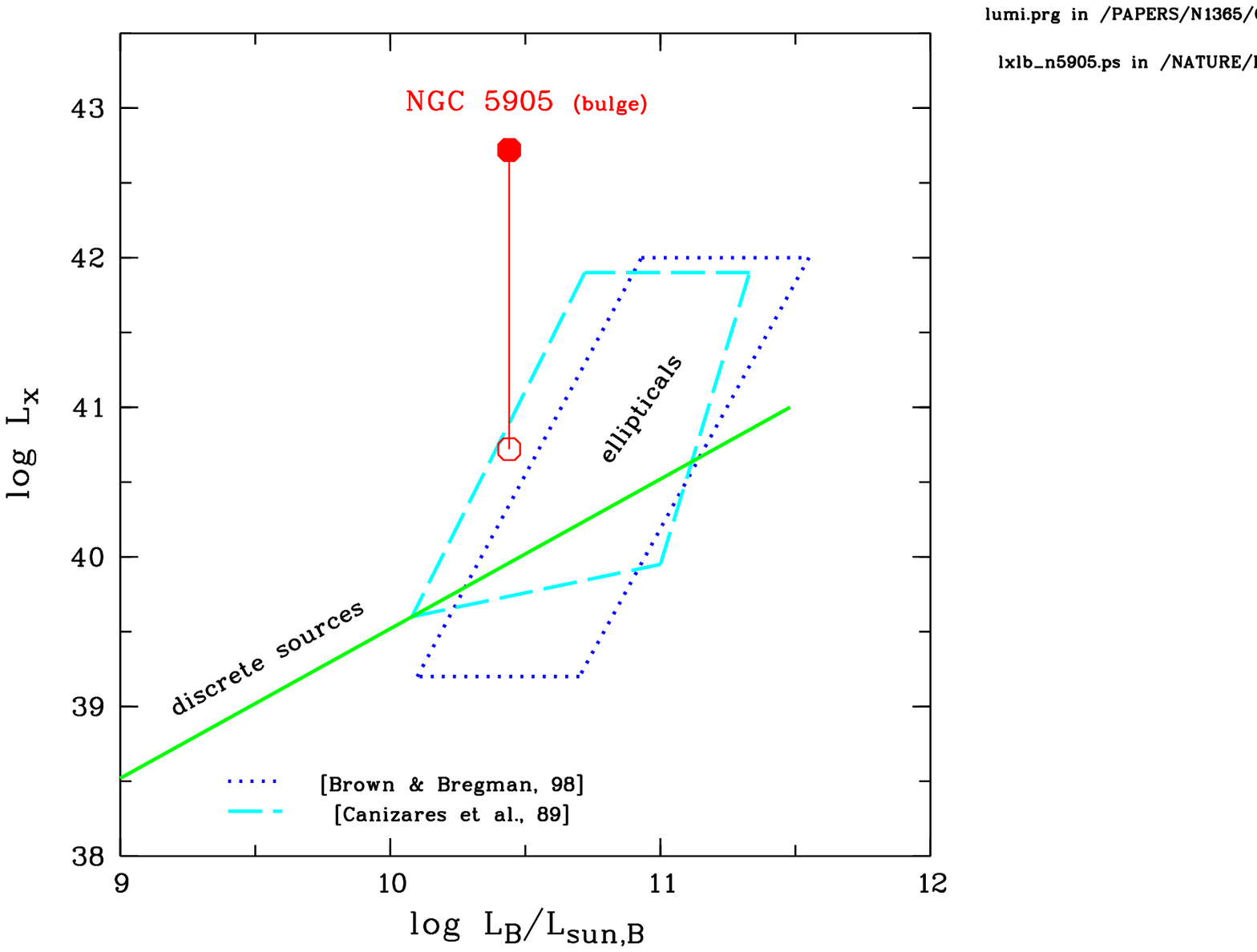,width=7.0cm,clip=} 
\hfill
\begin{minipage}[]{0.35\hsize}\vspace*{-6.7cm}
\hfill
\caption[lxlb]{ Position of NGC\,5905 in 
the $L_{\rm x} - L_{\rm B}$ diagram in outburst and low-state.
The dashed lines mark the region populated by
some samples of elliptical galaxies. 
}
\label{lxlb}
\end{minipage}
\end{figure}

\subsubsection{Properties of the host galaxies}

The expected rate of tidal disruption events is
about one event in {\em at least} $\sim$10$^4$ years per galaxy
(e.g., Magorrian \& Tremaine 1999), and the whole {\sl ROSAT}
data base has to be employed for a systematic search for
further tidal disruption events (for first results
see Komossa \& Bade 1999, and below).

The disruption rate depends on the efficiency with which the 
loss-cone orbits are 
re-filled
(Frank and Rees 1976, Lightman and Shapiro 1977, Shields \& Wheeler 1978).
This can be done by perturbations of the stellar orbits,
e.g. by merging events (Roos 1981) or a triaxial gravitational
potential in the galaxy's core (Norman \& Silk 1983).

Are the observed flaring galaxies special in this context
(i.e., did any process aid in re-populating the loss-cone orbits)?
So far, not much is known about the host galaxies of the few flaring
galaxies. The best studied case is NGC\,5905.
It is interesting to note that this galaxy 
posses multiple triaxial structures  
with a secondary bar (Friedli et al. 1996, Wozniak et al. 1995) which might
aid occasional tidal disruption events by disturbing the
stellar velocity fields. NGC\,5905 is in a pair with NGC\,5908.

RXJ1242$-$11 is actually a pair of galaxies at similar redshift, and
it is well possible that both galaxies are interacting.
The X-ray error circle of another outburster, RXJ1420+53, also
includes two galaxies. However, a redshift is so far only available
for the brighter of the two. 

The Hubble types of the flaring galaxies are not known in all cases.
NGC\,5905, of type SB (Keenan 1937), is one of the largest spiral galaxies known (e.g., Romanishin 1983),
whereas some of the other host galaxies look like ellipticals. 
Deeper optical imaging is presently in progress.

\section{\bf Search for further X-ray flares}

While we wait for the next-generation of X-ray all-sky surveys,
we can still make use of an existing data base
which has not yet been fully exploited, the {\sl ROSAT} data base
(Voges et al. 1999).

In a first step to search for further cases of strong X-ray variability
the sample of nearby galaxies of Ho et al. (1995) and
{\sl ROSAT} all-sky survey (RASS)
and archived pointed observations were used.
The sample of Ho et al. has the advantage of the availability
of optical spectra of good quality, which are necessary when searching
for `truly' non-active galaxies.
136 out of the 486 galaxies in the catalogue were detected
in pointed observations. The source countrates were then compared with those
measured during the RASS.

\subsection{Non-active galaxies}

No other X-ray flaring, optically in-active galaxy was found.
The absence of any further flaring event among the
sample galaxies is entirely consistent with the expected
tidal disruption rate of one event in $\sim$10$^{4-5}$ years per galaxy
(e.g., Magorrian \& Tremaine 1999).

The next step, presently in progress, will be an extended search for X-ray flaring
events based on the whole {\sl ROSAT} all-sky survey database.
This approach will allow statistical inferences on the abundance of SMBHs in
non-active galaxies (Sembay \& West 1993).

\subsection{AGN}

Several of the sample galaxies show variability by a factor 10--30. All
of these are well-known AGN.

Many active galactic nuclei are variable in X-rays
with a range of amplitudes, typically a factor 2--3
(e.g., Mushotzky et al. 1993, Ulrich et al. 1997).
The cause of variability is usually linked in
one way or another to the central engine; for instance by changes in
the accretion disk (e.g., Piro et al. 1988, 1997), or by variable obscuration
(e.g., Komossa \& Fink 1997a, Komossa \& Janek 2000).
None of the well-studied X-ray variable AGN of the present sample are candidates for
tidal disruption events since their lightcurves show
recurrent variability on different time scales. 

An example for a highly variable AGN among the present
sample galaxies is NGC\,4051.
Its long-term {\sl ROSAT} X-ray lightcurve 
exhibits variability in countrate by a factor $\sim$30.
Only a small part of the variability of NGC\,4051 can be explained
with a variable warm absorber,  the rest is likely intrinsic
(Komossa \& Fink 1997a, Komossa \& Janek 2001, and references therein).

Even higher total amplitude of variability is detected
in two subsequent {\sl ROSAT} observations of NGC\,3516. The X-ray
countrate varies by a factor $\sim$50 (Komossa \& Bade 1999,
Komossa \& Halpern 2001, in prep.); variable cold absorption
likely plays a major part in explaining the observations.


\section{\bf Future perspectives}

X-ray outbursts from non-active galaxies provide important information
on the presence of SMBHs in these galaxies,
and the link
between active and normal galaxies. 
One advantage of this
method compared to other approaches to
search for central dark masses -- like HST-based galaxy rotation curves -- 
is that the X-ray flare emission originates from the {\em very vicinity}
of the SMBH. Therefore, potentially it provides much tighter constraints 
on the black hole mass. Flares can also be detected out to larger cosmological distances.

\begin{figure}[h]
\hspace*{2.0cm}
\psfig{file=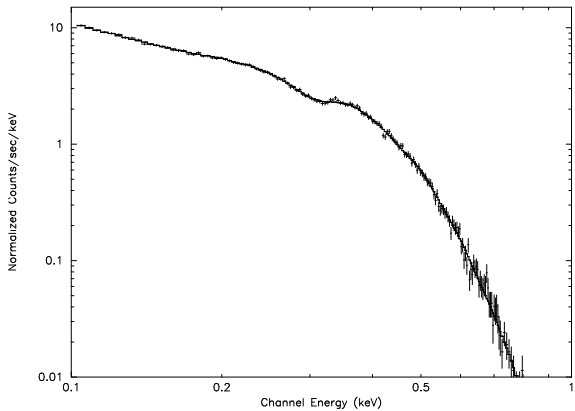,width=7.5cm,angle=-90,clip=}
\caption{Simulation of an XMM spectrum (EPIC pn instrument) of an X-ray flare similar to the
one observed from NGC\,5905, assuming a medium-deep exposure of 100 ksec and 
a single black body spectrum of $kT = 0.05$ keV absorbed
by $N_{\rm Gal} = 1.5\,10^{20}$ cm$^{-2}$ at a redshift of $z$=0.01.
At $z$=0.1   
the very soft black-body-like emission
is partly redshifted out of the instrument's
energy bandpass, causing a further drop in the observed countrate
($\sim$30\% at $z$=0.1) in addition to the flux decrease because
of the larger distance. The effect
increases with redshift such that very distant flares can no longer be
detected since shifted out of the observable energy range.
Among the X-ray flares, that of NGC\,5905 was of lowest luminosity. 
More luminous flares 
would therefore be detectable out to much higher redshifts.  
In addition to the continuum spectrum, we may expect to detect emission
features which arise in the accreted material, or absorption features from
the ISM of the host galaxy.  }
\label{xmm-simu}
 \end{figure}

\paragraph{Follow-up studies: future X-ray surveys.}
Future X-ray surveys,
like those planned with the {\sl LOBSTER} ISS X-ray all-sky monitor (Fraser 2001), 
{\sl MAXI} (Yuan et al. 2001, Mihara 2001),
and {\sl ROSITA} (Predehl 2001)
will be valuable in finding more of these outstanding
sources.
In addition, a number of flare events are expected to be detected 
(Yuan et al. 2002)
in pointings of the 
{\sl XMM-Newton} and {\sl Chandra} 
missions. 

On the one hand, all-sky surveys will be important in detecting 
the brightest events, due to
their large areal sky coverage.  Those surveys with sensitivity at the 
softest energies will be most efficient. 
On the other hand, deeper pointings on limited areas, or `pencil beam' surveys 
will increase the number of (more distant, on average fainter) events
(see Yuan et al. 2002 for logN-logS estimates of the expected number of flares
detectable with {\sl XMM-Newton}, and Sembay \& West 1993 for a general discussion). 
Concerning deep surveys in limited fields of view, two effects are
important. Firstly, there is a limiting maximal brightness the
events can reach in the context of the tidal disruption scenario:
the most luminous events are those with black hole masses around 10$^{8}$ M$_{\odot}$,
with high accretion rate, and the maximal possible fraction emitted in
the X-ray band (definite upper limit: $L_{\rm flare} \approxlt 10^{45-46}$ erg/s). 
Secondly, there is a limiting distance out to which flare
events are detectable because the flare spectra are very soft,
and for increasing redshift more and more of the (black-body-like) emission is shifted out
of the observable energy band (Fig. \ref{xmm-simu}).  In addition, distant 
galaxies may show a higher intrinsic fraction
of cold gas which will heavily absorb at soft X-rays.

After the discovery of new X-ray flare events,
{\sl rapid follow-up} multi-wave\-length observations 
will be essential.  
Apart from valuable new constraints on the favored outburst
scenario which would also allow a refinement of model calculations,  
these observations would enable us to 
address a number of important topics:   

\paragraph{Absorption-line spectroscopy of the IGM/ISM.}
As the flare emission travels through the ISM of the host galaxy,
and the IGM, absorption features will be imprinted on the X-ray spectrum.
These can then be used to study the properties of the absorbing material. 

\paragraph{Emission-line
spectroscopy of the circumnuclear material.}
If the soft X-ray flare emission has an extension into the EUV,
which is highly likely, then 
optical observations will
be important in order to detect potential emission lines
that were excited by the outburst emission.
Firstly, any gaseous material close to the nucleus is
expected to show an emission-line response.
The time variability of these lines will allow a
reverberation mapping of the circumnuclear gas;
line profiles, and line-ratios will allow
to estimate the velocity structure and physical conditions (density, abundances)
of this gas.  
In particular, this would also enable us to search for
the presence of a BLR in
these optically in-active galaxies.

\paragraph{Probing the realm of strong gravity.}
If observed with high spectral and temporal resolution
with the next generation of X-ray telescopes, like the {\sl XEUS} mission
sensitive between 0.05--30\,keV, 
the flare spectra may allow to probe the realm of strong gravity, since
the temporal evolution of the stellar debris,
and of potential spectral features, will depend
on relativistic precession effects around the Kerr metric.

\vskip0.8cm

\noindent {\sl Acknowledgements:}
It is a pleasure to thank 
the Astronomische Gesellschaft for awarding the 
Ludwig-Biermann price and for inviting me to give this talk.
I would like to thank 
Weimin Yuan, Michael Dahlem, Jules Halpern, Norbert Bade, Andrew Ulmer, Niel Brandt, 
Martin Elvis, David Meier, and the participants of the CAS-MPG Workshop on High-Energy Astrophysics
at Ringberg Castle 
for very useful discussions and comments on the subject of X-ray flares
from non-active galaxies, and Norbert Bade for taking the optical
image of NGC\,5905 which is shown in Fig. 4.
I gratefully remember Henner Fink for
introducing me to the work with X-ray data, for discussions
and helpful advice.
Henner Fink passed away in December 1996.
The {\sl ROSAT} project was supported by the German Bundes\-mini\-ste\-rium
f\"ur Bildung, Wissenschaft, Forschung und Technologie
(BMBF/DLR) and the Max-Planck-Society.
\\
Preprints of this and related papers can be retrieved at \\
http://www.xray.mpe.mpg.de/$\sim$skomossa/

\vspace{0.7cm}
\noindent
{\large{\bf References}}
{\small

\bref Antonucci R., 1993, 
    ARA\&A 31, 473 

\bref
 Ayal S., Livio M., Piran T., 2000, ApJ 545, 772

\bref
 Bade N., Komossa S., Dahlem M., 1996, A\&A 309, L35

\bref  Becker R.H., White R.L., Helfand D.J., 1995, ApJ, 450, 559

\bref  Beloborodov A.M., Illarionov A.F., Ivanov P.B., Polnarev A.G., 1992, MNRAS 259, 209

\bref  Bradt H., Levine A.M., Marshall F.E., et al. 2001, to appear in ESO Astrophysics Symposia,
F. Frontera et al. (eds), [astro-ph/0108004]

\bref  Brandt W.N., Pounds K.A., Fink H.H., 1995, MNRAS 273, L47

\bref Brandt W.N., Fabian A.C., Pounds K.A., 1996, MNRAS 278, 326 

\bref  Branduardi-Raymont G., Sako M., Kahn S., et al., 2001, A\&A 365, L140 

\bref  Brosch N., Krumm N., 1984, A\&A 132, 80

\bref Brown B.A., Bregman J.N., 1998, ApJ 495, L75 

\bref Canizares C.R., Fabbiano G., Trinchieri G., 1987, ApJ 312, 503 

\bref  Cannizzo J.K., Lee H.M., Goodman J., 1990, ApJ 351, 38

\bref  Carter B., Luminet J.P., 1985, MNRAS 212, 23

\bref Collin S., Abrassart A., Czerny B., Dumont A.-M., Mouchet M.,
           EDPS Conf. Series in Astron. \& Astrophys., in press  [astro-ph/0003108]

\bref  Collin-Souffrin S., Czerny B., Dumont A.-M., Zycki P.T., 1996, A\&A 314, 393

\bref  Collinge M.J., Brandt W.N.,
           Kaspi, S., et al., 2001, ApJ 557, 2

\bref  Condon J.J., Cotton W.D., Greissen E.W., et al., 1998, AJ 115, 1693

\bref Cui W., Feldkun D., Braun R., 1997, ApJ 477, 693 

\bref  Diener P., Frolov V.P., Khokhlov A.M., Novikov I.D.,
             Pethick C.J., 1997, ApJ 479, 164

\bref  Di\,Stefano R., Greiner R., Murray S., Garcia M., 2001, ApJ 551, L37 

\bref  Elvis M., 2000, ApJ 545, 63

\bref  Elvis M., 2001, in press,  [astro-ph/0109513]

\bref  Elvis M., Briel U., Henry J.P., 1983, ApJ 268, 105  

\bref  Elvis M., Fassnacht C., Wilson A.S., Briel U., 1990, ApJ 361, 459 

\bref  Eracleous M., Livio M., Binette L., 1995, ApJ 445, L1

\bref  Evans C.R., Kochanek C.S., 1989, ApJ 346, L13

\bref Fabian A., 2001, in: TEXAS Symposium, in press [astro-ph/0103438]

\bref Falcke H., 2001, in: Reviews in Modern Astronomy 14, R. Schielicke (ed.), 15 

\bref Ferland G.J., 1993, University of Kentucky, Physics Department, Internal Report

\bref  Ferrarese L., Merritt D., 2000, ApJ, 539, L9

\bref Ferrarese L., Pogge R.W., Peterson B.M., et al., 2001, ApJ 555, L79 

\bref  Franceschini A., Vercellone S., Fabian A.C., 1998, MNRAS 297, 817 

\bref  Frank J., Rees M.J., 1976, MNRAS 176, 633

\bref  Fraser G., 2001, in: MAXI workshop on AGN variability, in press

\bref  Friedli D., Wozniak  H., Rieke M., Martinet L.,
            Bratschi P., 1996, A\&AS 118, 461

\bref Gabriel A.H., Jordan C., 1969, MNRAS 145, 241  

\bref George I.M., Fabian A.C., 1991, MNRAS 249, 352  

\bref  Giuricin G., Bertotti G., Mardirossian F., Mezzetti M., 1990, MNRAS 247, 444

\bref Greenhill L.J., Henkel C., Becker R., Wilson T.L., Wouterloot J.G.A., 1995, A\&A 304, 21 

\bref  Greiner J., Schwarz R., Zharikov S., Orio M., 2000, A\&A 362, L25

\bref  Grupe D., Beuermann K., Mannheim K., et al.,
            1995, A\&A 299, L5

\bref  Grupe D., Leighly K., Thomas H., 1999, A\&A 351, L30

\bref  Gurzadyan V.G., Ozernoi L.M., 1979, Nature  280, 214

\bref  Gurzadyan V.G., Ozernoi L.M., 1980, A\&A 86, 315

\bref  Hills J.G., 1975, Nature  254, 295

\bref Halpern J.P., 1984, ApJ 281, 90

\bref  Ho L.C., Filippenko A.V., Sargent W.L.W., 1995, ApJS 98, 477

\bref Ho L.C., et al., 2001, ApJ, in press 

\bref  Hummel E., van der Hulst J.M., Keel W.C., Kennicutt R.C. Jr., 1987, A\&AS 70, 517

\bref Ikebe Y., Leighly K., Tanaka Y., et al., 2000,  MNRAS 316, 433 

\bref  Israel F.P., Mahoney M.J., 1990, ApJ 352, 30

\bref  Ivanov P.B., Novikov I.D., 2001, ApJ 549, 467

\bref Joseph R.D., 1999, Ap\&SS 266, 321

\bref Kaastra J., Mewe R., Liedahl D.A., Komossa S., Brinkman A.C., 2000, A\&A 354, L83

\bref Kaspi S., Brandt W.N., Netzer H., et al., 2001, ApJ 554, 216

\bref  Kato M., Hoshi R., 1978, Prog. Theor. Phys. 60/6, 1692

\bref Keenan P.C., 1937, ApJ 85, 325

\bref  Khokhlov A., Melia F., 1996, ApJ 457, L61

\bref  Kim S.S., Park M.-G., Lee H.M., 1999, ApJ 519, 647

\bref  Kochanek C., 1994, ApJ 422, 508  

\bref  Komossa S., 1999, in: ASCA/ROSAT Workshop on~AGN and the X-ray Background,
      T. Takahashi, H. Inoue (eds), ISAS Report, p. 149;
      [also available at astro-ph/0001263]

\bref Komossa S., 2001, A\&A 371, 507 

\bref  Komossa S., 2001, in: IX. Marcel Grossmann Meeting
 on General Relativity, Gravitation and Relativistic Field Theories,
 V. Gurzadyan et al. (eds), World Scientific, in press 

\bref  Komossa S., Bade N., 1998, A\&A 331, L49  

\bref  Komossa S., Bade N., 1999, A\&A 343, 775

\bref  Komossa S., B\"ohringer H., Huchra J., 1999, A\&A 349, 88

\bref  Komossa S., Dahlem M., 2001, in: MAXI workshop on AGN variability, ISAS Report, in press

\bref  Komossa S., Fink H., 1997a, A\&A 322, 719

\bref  Komossa S., Fink H., 1997b, A\&A 327, 483

\bref  Komossa S., Greiner J., 1999, A\&A 349, L45

\bref  Komossa S., Janek M., 2000, A\&A 354, 411

\bref Komossa S., Schulz H., Greiner J., 1998, A\&A 334, 110 

\bref Komossa S., Voges W., 2001, in: MPG-CAS workshop on high-energy astrophysics, 
        preprint available at: http://www.xray.mpe.mpg.de/~skomossa/publrev.html 

\bref Koratkar A., Deustua S.E., Heckman T., et al., 1995, ApJ 440, 132

\bref  Kormendy J., Richstone D.O., 1995, ARA\&A 33, 581

\bref  Kormendy J., Gebhardt K., 2001, in press [astro-ph/0105230]

\bref Krolik J.H., Kriss G.A., 1995, ApJ 447, 512 

\bref  Laguna P., Miller W.A., Zurek W.H., Davies M.B., 1993, ApJ 410,
              L83

\bref  Lee H.M., Kang H., Ryu D., 1995, ApJ 464, 131

\bref  Lee J., Ogle P., Canizares C., et al., 2001, ApJ 554, L13

\bref Lee J., 2001, talk given at: Workshop on X-ray spectroscopy of active galactic nuclei 
                with Chandra and XMM-Newton  (Garching, Dec. 2001) 

\bref  Lidskii V.V., Ozernoi L.M., 1979, Sov. Astron. Lett. 5(1), 16

\bref Liedahl D.A., Kahn S.M., Osterheld A.L., Goldstein W.H., 1990, ApJ 350, L37   

\bref  Lightman A.P., Shapiro S.L., 1977, ApJ 211, 244

\bref  Loeb A., Ulmer A., 1997, ApJ 489, 573

\bref  Luminet J.P., Marck J.-A., 1985, MNRAS 212, 57

\bref  Lynden-Bell D., 1969, Nature  223, 690

\bref  Magorrian J., Tremaine S., 1999, MNRAS 309, 447

\bref  Matsumoto C., Inoue H., 2001, in: MAXI workshop on AGN variability, in press

\bref  Matt G., Perola G.C.,
         Fiore F., et al., 2000, A\&A 363, 863   

\bref  Mihara T., 2001, in: MAXI workshop on AGN variability, in press

\bref Mitsuda K., 1995, Ann.N.Y.Acad.Sc. 759,
         Proc. 17th Texas Symp. Relat. Ap. and Cosm.,
 H. B{\"o}hringer, G.E. Morfill, J.E. Tr{\"u}mper (eds), 213

\bref Miyoshi M., Moran J., Herrnstein J., et al., 1995, Nature 373, 127

\bref Morales R., Fabian A.C., 2001, MNRAS, in press [astro-ph/0109050]

\bref  Mushotzky R., 1982, ApJ 256, 92 

\bref  Mushotzky R.F., Done C., Pounds K.A., 1993, ARA\&A 31, 717

\bref Nakagawa T., Kii T., Fujimoto R., et al., 1999, Ap\&SS 266, 43 

\bref  Nandra P., 2001, talk given at: Workshop on X-ray spectroscopy of active galactic nuclei
                with Chandra and XMM-Newton  (Garching, Dec. 2001)

\bref  Nandra K., Pounds K.A., 1992, Nature 359, 215 

\bref Netzer H., 1993, ApJ 411, 594 

\bref Neufeld D.A., Maloney P.R., 1995, ApJ 447, L17

\bref  Nolthenius R.A., Katz J.I, 1982, ApJ 263, 377

\bref  Nolthenius R.A., Katz J.I, 1983, ApJ 269, 297

\bref Norman C., Silk J., 1983, ApJ 266, 502  

\bref Ogle P.M., Marshall H.L., Lee J.C.,
             Canizares C.R., 2000, ApJ 545,  L81 

\bref  Peterson B.M., Ferland G.J., 1986, Nature  324, 345

\bref  Peterson B.M., 2001, in: Advanced lectures on the starburst-AGN connection,
I. Aretxaga et al. (eds), World Scientific, 3

\bref  Peterson B.M., Wandel A., 2000, ApJ 540, L13

\bref  Piro L., Massaro E., Perola G.C., Molteni D., 1988, ApJ 325, L25

\bref  Piro L., Balucinska-Church M., Fink H.,
                et al., 1997, A\&A 319, 74

\bref  Predehl P., 2001, in: MAXI workshop on AGN variability, in press

\bref Ptak A., Yaqoob T., 
 Mushotzky R.,
 Serlemitsos P., Griffiths R., 1998, 501, L37 

\bref  Ptak A., Serlemitsos P., Yaqoob T., Mushotzky R., 1999, ApJS 120, 179

\bref  Rauch K.P., 1999, ApJ 514, 725

\bref  Rauch K.P., Ingalls B., 1998, MNRAS 299, 1231

\bref  Rees M.J., 1988, Nature  333, 523

\bref  Rees M.J., 1989, Rev. mod. Astr. 2, R. Schielicke (ed.), 1
 
\bref  Rees M.J., 1990, Science 247, 817   

\bref  Rees M.J., 1997, Rev. mod. Astr. 10, R. Schielicke (ed.), 179

\bref  Reiprich T., Greiner J., 2001, in:
          ESO workshop on black holes in binaries and AGN, p. 168

\bref  Renzini A., 
 Greggio L., Di Serego Alighieri S., et al.,
           1995, Nature  378, 39

\bref Reynolds C.S., Ward M.J., Fabian A.C., Celotti A., 1997, MNRAS 291, 493 

\bref Rigopoulou D., Lawrence A., Rowan-Robinson M., 1996, MNRAS 278, 1049

\bref Roberts T.P., Warwick R.S., Ohashi T., 1999, MNRAS 304, 52 

\bref  Romanishin W., 1983, MNRAS 204, 909

\bref  Roos N., 1981, A\&A 104, 218

\bref  Roos N., 1992, ApJ 385, 108

\bref  Sadler E., Slee O.B., Reynolds J.E., Roy A.L., 1995, MNRAS 276, 1373

\bref Sako M., Kahn S.M., Behar E., et al., 2001, A\&A 365, L168  

\bref Sanders D.B., 1999, Ap\&SS 266, 331 

\bref Sanders D.B., Mirabel I.F., 1996, ARA\&A 34, 749 

\bref  Sanders R.H., 1984, A\&A 140, 52

\bref  Sanders R.H., van Oosterom W., 1984, A\&A 131, 267  

\bref  Salucci P., Ratnam C., Monaco P., Danese L., 2000, MNRAS 317, 488

\bref Schartel N., Komossa S., Brinkmann W., Fink H.H., Tr\"umper J., 
 Wamsteker W., 1997, A\&A 320, 421

\bref Schulz H., Komossa S., 1999, contribution to: Bad Honnef workshop on
             Trends in Astrophysics and Cosmology, astro-ph/9905118 

\bref Schulz H., Komossa S., Bergh\"ofer T., Boer B., 1998, A\&A 330, 823

\bref  Sembay S., West R.G., 1993, MNRAS 262, 141

\bref  Shields G.A., Wheeler J.C., 1978, ApJ 222, 667

\bref  Staveley-Smith L., Davies R.D., 1987, MNRAS 224, 953

\bref Svensson R., 1994, ApJS 92, 585 

\bref  Syer D., Ulmer A., 1999, MNRAS 306, 35

\bref  Tanaka  Y., Nandra K., Fabian A.C., et al., 1995, Nature  375, 659

\bref  Ulmer A., Paczynski B., Goodman J., 1998, A\&A 333, 379

\bref 
 Ulrich, M.-H., 
 Maraschi L., Urry C.M., 1997, ARA\&A 35, 445

\bref  van Moorsel G.A., 1982, A\&A 107, 66

\bref Vignati P., Molendi S., Matt G., et al., 1999, A\&A 349, L57 

\bref  Voges W., et al., 1999, A\&A 349, 389

\bref Wilman R.J., 1999, Ap\&SS 266, 55

\bref Wilms J., Reynolds C., Begelman M.C., et al., 2001, MNRAS\,L, in press [astro-ph/0110520]

\bref  Wozniak H., Friedli D., Martinet L., Martin P., Bratschi P., 1995, ApJS 111, 115

\bref Wu X.-B., Han J.L., 2001, A\&A 380, 31 

\bref Yaqoob T., George I.M., Turner T.J., 2002, ASP conf. series, in press [astro-ph/0111428] 

\bref Yi I., Boughn S.P., 1998, ApJ 499, 198 

\bref Yi I., Boughn S.P., 1999, ApJ 515, 579 

\bref  Young P.,
                     Shields G.,
           Wheeler J.C., 1977, ApJ 212, 367

\bref  Yuan W., Matsuoka M., Shirasaki Y., et al., 2001, in: MAXI workshop on AGN variability, in press 

\bref Yuan W., et al., 2002, in: New Visions of the X-ray Universe,  F. Jansen et al. (eds.), in press

}

\vfill

\end{document}